\documentclass[3p,times]{elsarticle}

\usepackage{ecrc}

\volume{00}

\firstpage{1}

\journalname{Nuclear Physics A}

\runauth{P. Gubler et al.}

\jid{NUPHA}

\jnltitlelogo{Nuclear Physics A}

\CopyrightLine{2012}{Published by Elsevier Ltd.}

\usepackage{amssymb}

\usepackage[figuresright]{rotating}

\begin{document}

\begin{frontmatter}

\dochead{}



\title{Modification of hadronic spectral functions under
extreme conditions: An approach based on QCD sum
rules and the maximum entropy method}

\author[a]{Philipp Gubler}\author[b]{Kei Suzuki}\author[c]{Kenji Morita}\author[b,d]{Makoto Oka}

\address[a]{RIKEN Nishina Center, Hirosawa 2-1, Wako, Saitama, 351-0198, Japan}

\address[b]{Tokyo Institute of Technology, Meguro 2-12-1, Tokyo 152-8551, Japan}

\address[c]{Yukawa Institute for Theoretical Physics, Kyoto University, Kyoto 606-8502, Japan}

\address[d]{J-PARC Branch, KEK Theory Center, Institute of Particle and Nuclear Studies, 203-1, Shirakata, Tokai, Ibaraki, 319-1106, Japan}

\begin{abstract}
Studies of 
quarkonium spectral functions at finite temperature, based on an approach combining 
QCD sum rules and the maximum entropy method are briefly reviewed. 
QCD sum rules for heavy quarkonia incorporate finite temperature effects in form of changing values of gluonic condensates 
that appear in the operator product expansion. These changes depend on the energy density and 
pressure at finite temperature, which we extract from quenched lattice QCD calculations. 
The maximum entropy method then allows us to obtain the most probable spectral function from 
the sum rules, without having to introduce any 
specific assumption about its functional form. 
Our findings suggest that the charmonium ground states of both S-wave and P-wave channels dissolve into the continuum 
already at temperatures around or slightly above the critical temperature $T_c$, while the bottomonium states 
are less influenced by temperature effects, surviving up to about 2.5 $T_c$ or higher for S-wave and up to about 2.0 $T_c$ for 
P-wave states. 
\end{abstract}

\begin{keyword}

Quarkonium \sep QCD sum rules \sep QCD at finite temperature

\end{keyword}

\end{frontmatter}


\section{Introduction}
\label{Intro}
The study of quarkonia in hot matter has, since the early suggestions \cite{Matsui} that were made more than 25 years ago, evolved into 
a field with diverse activities in both experiment and theory. Especially, through the heavy-ion collision experiments at RHIC and LHC, a huge amount 
of experimental data on quarkonium production in various reactions is now available, which can be compared with theoretical expectations. 
This task, however, has turned out to be very complex, as a large number of competing effects have to be taken into account to describe the 
experimental results (for reviews see \cite{Rapp,Kluberg}). The most basic inputs of these calculations are the 
quarkonia spectral functions, which include all the physically relevant information of the quarkonium states as well as their behavior 
at finite temperature. 
It is therefore desirable to obtain these spectral functions from calculations based on QCD. Progress in this direction has been acheived 
mostly due to the advancement of lattice QCD and the use of the maximum entropy method (MEM) for the extraction of the spectral function 
from the euclidean time correlator \cite{Asakawa} (see also \cite{Ding} and the references therein for the latest results). 

Another approach for capturing information on quarkonia spectral functions at both zero and finite temperature is provided by QCD sum rules. 
This method exploits the analytic properties of the two-point function of operators to 
derive certain integrals over the hadronic spectral functions (the ``sum rules"), which, via the operator product expansion (OPE), 
can be related to a combination of perturbatively calculable 
quantities and non-perturbative condensates, containing information on the QCD vacuum. In the case of the quarkonia 
channels considered here, these are gluonic condensates, the most important one being the gluon condensate of mass dimension 4 \cite{Morita1}. 

Recently, it has become possible to make use of the MEM technique to analyze QCD sum rules \cite{Gubler1}, which allows 
to extract the most probable form of the spectral function from the OPE without having to resort to some specific 
functional form. 
This approach has since been applied to both charmonium \cite{Gubler2} and bottomonium \cite{Suzuki} channels and to 
discuss these results will be main goal of this article. 

The proceedings are organized as follows. After a brief description of our formalism and analysis methods, we recapitulate 
the results obtained so far in 
\cite{Gubler2,Suzuki}. We will also show some novel results on the thermal behavior P-wave charmonia ($\chi_{c0}$, $\chi_{c1}$), 
which were not included in \cite{Gubler2}. 
Finally, we summarize our results and give an outlook by 
discussing open issues and possible future directions of further improvements of the sum rule approach presented here. 

\section{Formalism}
\label{Form}
QCD sum rules at finite temperature make use of the 
analyticity of the two-point correlator of some general local operator $j^{\mathrm{J}}(x)$:
\begin{equation} 
\Pi^{\mathrm{J}}(q) = i \displaystyle \int d^4x e^{iqx} 
\langle T [j^{\mathrm{J}}(x) j^{\mathrm{J}}(0) ] \rangle_T. 
\label{eq:correlator} 
\end{equation}
Here, $j^{\mathrm{J}}(x)$ 
stands for $\bar{h} \gamma_{\mu} h(x)$, $\bar{h} \gamma_{5} h(x)$, 
$\bar{h} h(x)$ and $(q_{\mu} q_{\nu} /q^2 - g_{\mu\nu})\bar{h} \gamma_5 \gamma^{\nu} h(x)$ in the 
vector, pseudoscalar, scalar and axial-vector channel, respectively, while 
$h(x)$ represents either a charm- or bottom-quark field. 
The definition of the expectation value $\langle \mathcal{O} \rangle_T$ 
is $\langle \mathcal{O} \rangle_T \equiv \mathrm{Tr}( e^{-H/T} \mathcal{O} ) / \mathrm{Tr}( e^{-H/T} )$. 
Through a dispersion relation, one can relate the correlator 
calculated in the deep-Euclidean region ($-q^2 \to \infty$) to a specific integral over the hadronic 
spectral function $\rho^{\mathrm{J}}(s)$. After the application of the Borel transform, one 
arrives at the following expression:
\begin{equation}
\mathcal{M}^{\mathrm{J}}(M^2) = \displaystyle \int_0^{\infty} ds 
e^{-s/M^2} \rho^{\mathrm{J}}(s). 
\label{eq:disp}
\end{equation}
In this equation, the left-hand side can be calculated analytically using 
the OPE. To get the spectral function $\rho^{\mathrm{J}}(s)$ 
one therefore somehow has to invert the above integral. This is, 
however, an ill-posed problem and can not be solved rigorously, because the left-hand side 
is only known as an asymptotic expansion, with coefficients determined 
with limited precision. 
With the help of Bayes' theorem, it is nevertheless possible to obtain the most probable 
form of $\rho^{\mathrm{J}}(s)$, given Eq. (\ref{eq:disp}) and additional 
information on the spectral function such as positivity and 
asymptotic values at high and low energy. This is the essence of the MEM approach discussed in this article. For the 
concrete implementation of MEM for analyzing QCD sum rules, see \cite{Gubler1}. 

For the sum rules of quarkonia, all the finite temperature effects can be 
included into the temperature dependent condensates owing to the large separation scale 
in the OPE \cite{Hatsuda}. 
This speration is valid as long as the temperature does not reach values of the order of the 
quark mass. 
The temperature dependencies of the 
condensates have been obtained in \cite{Morita1}, where the relation between the gluon 
condensate and the energy momentum tensor is exploited to give the value of the 
gluon condensate as a function of the energy 
density and pressure. These thermodynamic quantities are then extracted from 
quenched lattice QCD calculations \cite{Boyd,Kaczmarek}. 
In this way, we are able to calculate the left-hand side of the sum rules as a function of 
temperature. As a last step, we then use MEM to retrieve the corresponding spectral 
functions from Eq. (\ref{eq:disp}). 

\section{Results}
\label{Res}
\begin{figure}[t]
\begin{center}
\includegraphics[width=0.40\textwidth]{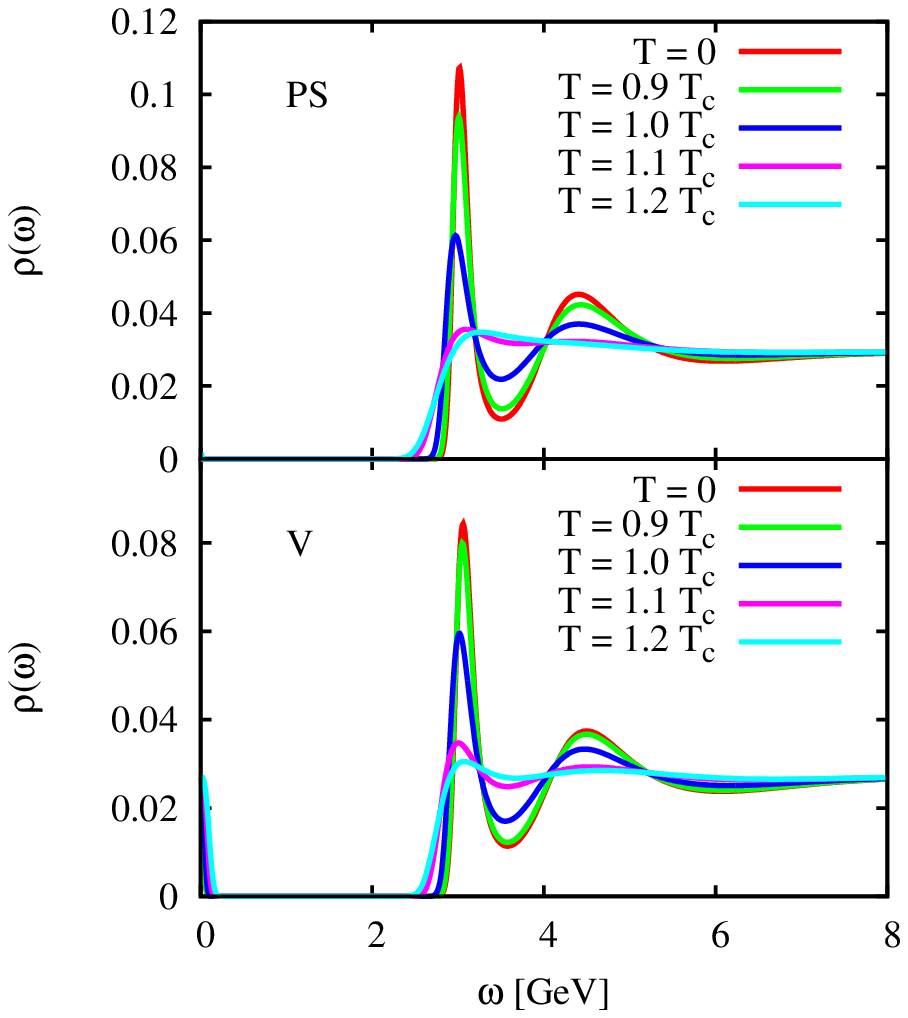}
\hspace{1.0cm}
\includegraphics[width=0.40\textwidth]{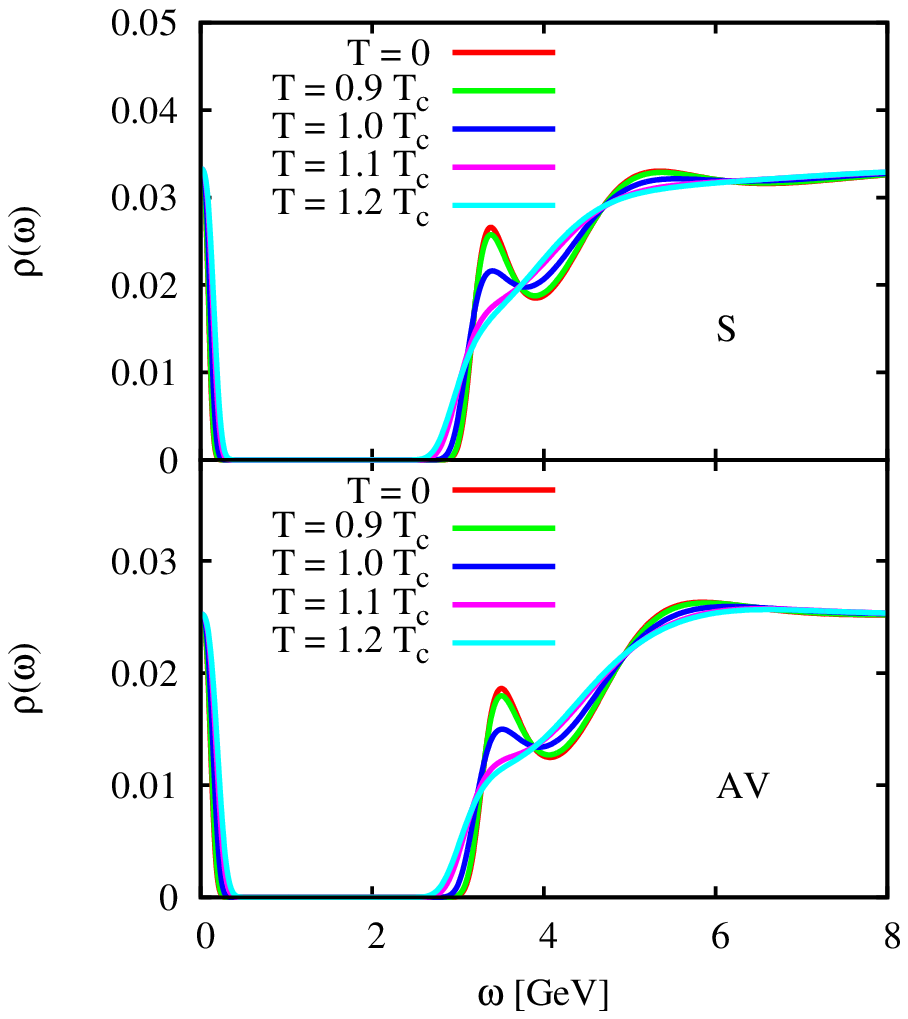}
\caption{Charmonium spectral functions at zero and finite temperature in the pseudoscalar (top left), sclalar (top right), 
vector (bottom left) and axialvector channel (bottom right). The left figures are adapted from 
\cite{Gubler2}.}
\label{fig:Fig1}
\end{center}
\end{figure}

Let us first discuss the results of the charmonium channels. They are shown in Fig. \ref{fig:Fig1}. 
The S-wave channels (left plots) have already been presented in \cite{Gubler2}, while the 
P-wave channels are shown here for the first time. 
Concentrating firstly on the spectral functions at zero temperature, it is seen that clear peaks 
are generated, which represent the lowest state of each channel. The positions of these peaks reproduce the experimental 
values with a precision of about $50\,\mathrm{MeV}$. 

At finite temperature, we observe that 
the lowest peaks of all channels vanish 
slighly above the critical temperature $T_c$. 
The origin of this melting effect is a sudden 
change of the gluonic condensates around $T_c$, which can be 
related to 
the deconfinement transition of the gluonic matter. 

\begin{figure}[t]
\begin{center}
\includegraphics[width=0.40\textwidth]{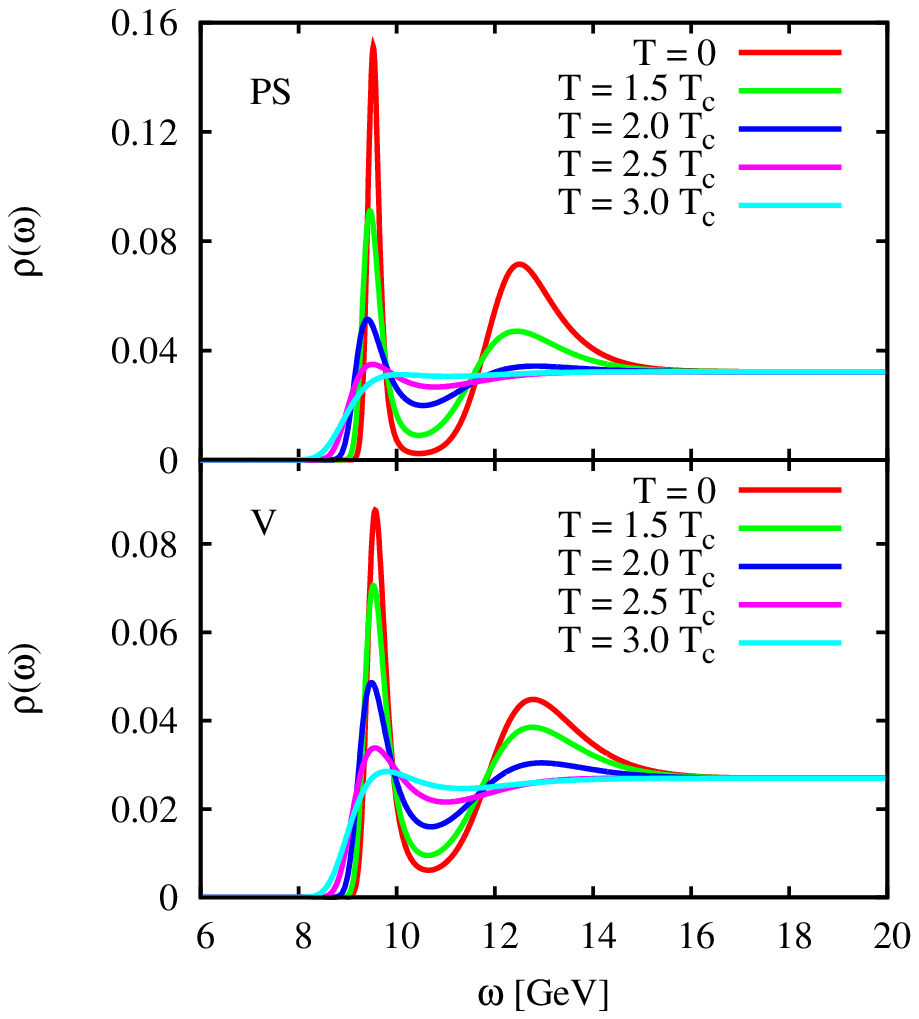}
\hspace{1.0cm}
\includegraphics[width=0.40\textwidth]{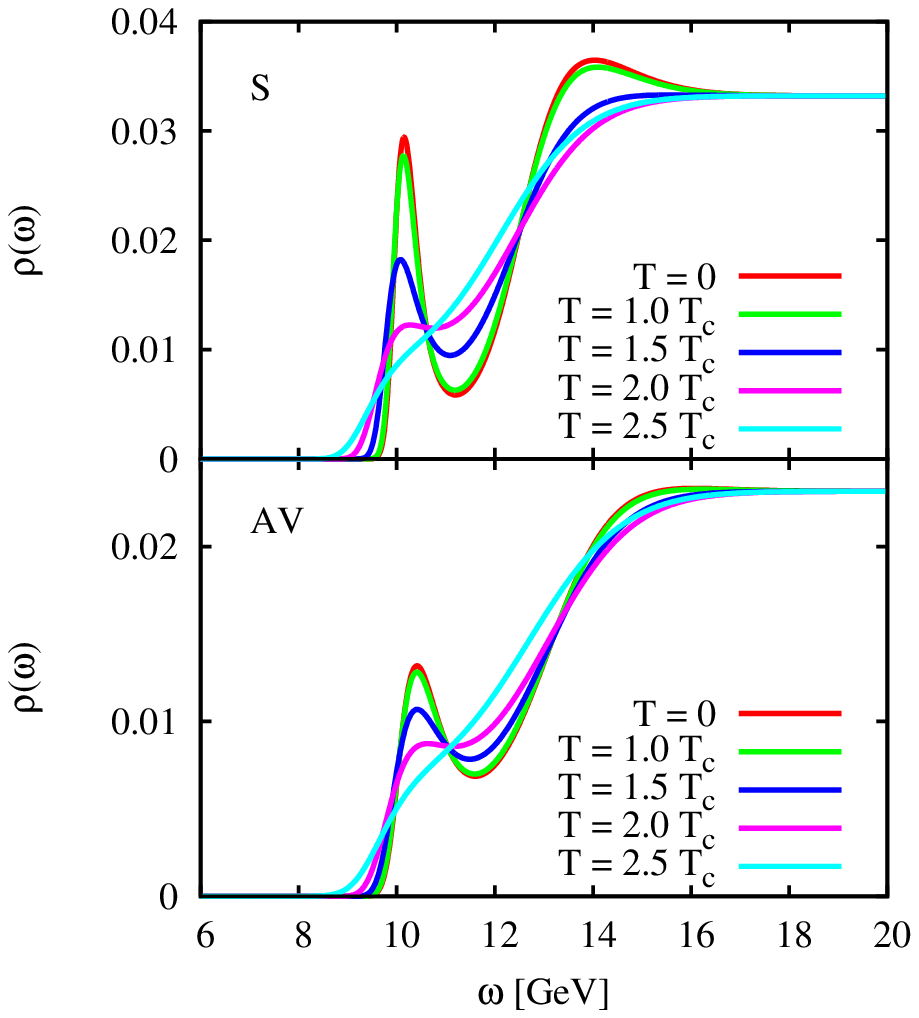}
\caption{Bottomonium spectral functions at zero and finite temperature in the pseudoscalar (top left), sclalar (top right), 
vector (bottom left) and axialvector channel (bottom right). All shown figures are adapted from 
\cite{Suzuki}.}
\label{fig:Fig2}
\end{center}
\end{figure}

Next, let us look at the results for bottomonium. These are given in Fig. \ref{fig:Fig2}. 
Here, as for charmonium, clear peaks are seen for each channel at zero temperature. 
These peaks are found at 150-500 MeV above the experimental values of the respective ground states. This discrepancy is caused 
by the excited states (for instance $\Upsilon(2S)$ and $\Upsilon(3S)$ in the vector channel), which cannot be resolved by the MEM 
analysis, and pull the lowest peaks to higher energies than the actual ground state. This is in contrast to the 
charmonium case, in which the excited states give only a relatively small contribution to the lowest peak. 
For a more detailed discussion of this issue, see \cite{Suzuki}. 

Turning now to the finite temperature curves, it is noted that the bottomonium states are modified much slower than their charmonium 
counterparts, which is in agreement with phenomenological expectations. 
Concretely, both S-wave spectral functions still exhibit a clear peak at $T=2.0\,T_c$ which starts to dissolve at about $2.5\,T_c$. 
The P-wave states are on the other hand modified somewhat faster and disappear already at temperatures around $2.0\,T_c$. The reason 
for the robustness of the bottomonium states can be traced back to the fact that the gluon condensate terms in the OPE 
are proportional to $1/m^4_h$, $m_h$ being the quark mass. These are the driving terms of the quarkonium melting, and are therefore 
relatively suppressed for the bottomonium sum rules. This is why for heavier quark masses one needs to go to higher temperatures to observe a 
significant effect. 

\section{Conclusions and Outlook}
\label{DiaOu}
We have analyzed quarkonia ground states at finite temperature in the vector-, 
pseudoscalar-, scalar- and axialvector-channels. 
By combining the techniques of QCD sum rules and MEM, we have extracted the spectral functions 
from the OPE of the correlators calculated in the deep Euclidean region. 
As a result, it is found that the charmonium ground states of both S-wave and P-wave channels dissolve into the continuum 
already at temperatures around $T_c$, while the bottomonium states 
are more stable, surviving up to about 2.5 $T_c$ or higher for S-wave and about 2.0 $T_c$ for 
P-wave states. 

On a qualitative level, 
the results shown in the previous section appear to be quite reasonable. 
We, however, have assumed that the OPE converges 
sufficiently fast and thus the higher order terms will consist of only small corrections by which 
the general picture is not altered. 
The only way of checking the validity of this assumption, is to take into account more higher order terms and to evaluate their 
effect on the spectral functions. 
The OPE 
generally contains two sorts of terms: perturbative and non-perturbative. The perturbative terms include the higher 
order $\alpha_s$ corrections to the Wilson coefficients appearing in the OPE. Among them, especially the second order correction to 
the identity operator Wilson coefficient is potentally large \cite{Ioffe}, and a detailed calculation of its contribution to the sum rules of the 
vector channel is presently ongoing \cite{Gubler3}. Furthermore, the non-perturbative corrections contain various gluonic 
condensates of higher dimension, for whose one ideally needs to know the value at zero and finite temperature. Such information 
can only be accessed via a lattice QCD calculation, which is planned to be carried out in the future. 

\section*{Acknowledgments}
This work is supported by KAKENHI under Contract Nos. 22105503, 
19540275, 24540271 and 
by YIPQS at the Yukawa Institute for Theoretical Physics. 
P.G. gratefully 
acknowledges the support by the Japan Society for the Promotion of Science for Young 
Scientists (Contract No. 21.8079). 






\begin{thebibliography}{00}


\bibitem{Matsui} 
T. Matsui and H. Satz, 
Phys. Lett. B \textbf{178}, 416 (1986).
\bibitem{Rapp}
R. Rapp, D. Blaschke, and P. Crochet, 
Prog. Part. Nucl. Phys. \textbf{65}, 209 (2010).
\bibitem{Kluberg}
L. Kluberg and H. Satz, 
arXiv:0901.3831 [hep-ph]. 
\bibitem{Asakawa}
M. Asakawa and T. Hatsuda, 
Phys. Rev. Lett. \textbf{92}, 012001 (2004).
\bibitem{Morita1}
K. Morita and S.H. Lee, 
Phys. Rev. Lett. \textbf{100}, 022301 (2008).
\bibitem{Ding}
H.T. Ding \textit{et al.}, 
Phys. Rev. D \textbf{86}, 014509 (2012). 
\bibitem{Gubler1}
P. Gubler and M. Oka, 
Prog. Theor. Phys. \textbf{124}, 995 (2010). 
\bibitem{Gubler2}
P. Gubler, K. Morita and M. Oka, 
Phys. Rev. Lett. \textbf{107}, 092003 (2011). 
\bibitem{Suzuki}
K. Suzuki, P. Gubler, K. Morita and M. Oka, 
Nucl. Phys. \textbf{A897}, 28 (2013). 
\bibitem{Hatsuda}
T. Hatsuda, Y. Koike and S.H. Lee, 
Nucl. Phys. \textbf{B394}, 221 (1993). 
\bibitem{Boyd}
G. Boyd \textit{et al.}, 
Nucl. Phys. \textbf{B469}, 419 (1996).
\bibitem{Kaczmarek}
O. Kaczmarek \textit{et al.}, 
Phys. Rev. D \textbf{70}, 074505 (2004). 
\bibitem{Ioffe}
B.L. Ioffe and K.N. Zyablyuk, 
Eur. Phys. J. C \textbf{27}, 229 (2003).
\bibitem{Gubler3}
P. Gubler, K. Suzuki, K. Morita and M. Oka, 
in  progress. 

\end{thebibliography}



\end{document}